\documentclass{aa}  
\usepackage{graphicx}
\usepackage{txfonts}
\begin{document}
   \title{Spin-up and hot spots can drive mass out of a binary}

   \author{W. Van Rensbergen, J.P. De Greve, C. De Loore, \and N. Mennekens}

   \offprints{W. Van Rensbergen}

   \institute{Astrophysical Institute, Vrije Universiteit Brussel, Pleinlaan 2, 1050 Brussels, Belgium\\
   \email {wvanrens@vub.ac.be}
   }

    \date{Received April 08, 2008; accepted June 26, 2008}

 
  \abstract
   {The observed distribution of orbital periods of Algols with a B-type primary at birth agrees fairly well with  the prediction from conservative theory. Conservative evolution fails, however, to produce the rather large fraction of Algols observed with a high mass-ratio, especially: $q$ $\in$ [0.4-0.6].}
   {In order to keep Algols for a longer time with a higher mass-ratio without disturbing the distribution of orbital periods too much, interacting binaries  have to lose a significant fraction of their total mass without losing much angular momentum before or during $Algolism$. We propose a mechanism that meets both requirements.}
   {In the case of direct impact the gainer spins up: sometimes up to critical velocity. Equatorial material on the gainer is therefore less bound. A similar statement applies to material located at the edge of an accretion disc. The incoming material moreover creates a hot spot in the area of impact. The sum of the rotational and radiative energy of hot spot material depends on the mass-transfer-rate. The sum of both energies overcomes the binding energy at a well defined critical value of the mass-transfer-rate. As long as the transfer-rate is smaller than this critical value RLOF happens $conservatively$. But as soon as the critical rate is exceeded the gainer will acquire no more than the critical value and RLOF runs into a $liberal$ era.}
   {Low-mass binaries never achieve mass-transfer-rates larger than the critical value. Intermediate-mass binaries evolve mainly conservatively but mass will be blown away from the system during the short era of rapid mass-transfer soon after the onset of RLOF. We have calculated the evolution of binaries with a 9  $M_{\odot}$ primary and a 5.4 $M_{\odot}$ companion over a range of initial orbital periods, covering case-A RLOF. Mass-loss from the system is achieved during direct impact only.}
   {We find systems that show $Algolism$ for more than ten million years. RLOF occurs almost always $conservatively$. Only during some 20,000 years the gainer is not capable of grasping all the material that comes from the donor.  The mass-ratio $q$ $\in$ [0.4-0.6] which was hardly populated by $conservative$ evolution now contains Algols for a significant fraction of their existence.}

   \keywords{binaries: eclipsing - stars: evolution - stars: mass-loss - stars: statistics}
   
    \authorrunning{W. Van Rensbergen et al.}  
     \titlerunning{Spin-up and hot spots can drive mass out of a binary}
   \maketitle

\section{Introduction}

Eggleton (\cite{Eggleton}) introduced the denomination $liberal$ to distinguish binary evolution with mass and subsequent angular momentum loss from the $conservative$ case, where no mass leaves the system.  Liberal evolution must be at work since Refsdal et al. (\cite{Refsdal}) found no progenitor that can evolve into AS Eri in a $conservative$ way. Massevitch  \& Yungelson  (\cite{Massevitch}) showed that in order to obtain agreement between theory and observations systems with combined mass $\le$ 6 $M_{\odot}$ have to lose 40 to 50 $\%$ of the mass lost by the donor. Sarna (\cite{Sarna}) showed that only 60 $\%$ of the mass lost by the loser of $\beta$ Per was captured by the gainer, while 30 $\%$ of the angular momentum was lost during Roche Lobe Overflow (RLOF). It has been shown by the $Brussels$  group that $conservative$ calculations produce almost no Algols with large mass-ratios during case-B RLOF (Van Rensbergen, \cite{Walter}). Van Rensbergen et al. (\cite{Walter1})  included case-A RLOF into the comparison between observation and $conservative$ binary evolution theory, leading again to a too small number of Algols with large mass-ratios. The observed distribution of orbital periods of Algols is, however, fairly well reproduced by $conservative$ evolution. All the conservative evolutionary tracks can be found at  \verb"http://www.vub.ac.be/astrofys/". In this paper we propose a scenario wherein violent phases of rapid RLOF can trigger mass-loss from the system. Whereas Algols at present go mainly through quiet phases of RLOF, they may have had a violent past in which they have lost a considerable fraction of their mass.
      

\section{Observed orbital periods and mass-ratios}

%
To avoid an eternal confusion we replace the indices (1= $primary$ and 2=$secondary$) characterizing binary components by $d$ and $g$. We use the index $d$  for the $donor$, i.e. the star that will be the donor once RLOF has started, whereas $g$ is used for its gaining companion. 
We define the mass-ratio $q$ throughout as:

 \begin{equation}
      q = {M_{d}\over M_{g}} 
\label{eqn:qratio}
   \end{equation}
 
   De Loore   \& Van Rensbergen (\cite{Bert}) introduced the qualification $Algolism$ for binaries during their Algol stage, when the mass of the $donor$ has become necessarily smaller than the mass of the $gainer$. In our comparative study the value of the mass-ratio  $q$ will thus always be in the interval [0-1].
   
   \vspace{0.2cm}
   
   In this paper we compare theory with observations for Algol binaries with a B-type primary at birth. The catalogue of Budding et al. (\cite{Budding et al.}) extended with the semi-detached Algols from Brancewicz et al. (\cite{Brancewicz et al.}) supplies us with 303 Algols which can be issued from $conservative$ binary evolution with a B-type primary at birth.
   
   The observed distribution of orbital periods and mass-ratios for the SB2s among these systems is well established. The overall distribution of mass-ratios also includes the SB1s and has been revisited since Van Rensbergen  et al. (\cite{Walter1}) claimed that more than 70$\%$ of the observed Algols are located in  $q$ $\in$ [0.4-1] if one uses $q_{MS}$ which is determined so as to make the parameters of the most massive star fit main sequence characteristics. Two other methods evaluating mass-ratios of SB1s were, however, not used in this study: the mass-ratio $q_{LC}$ which is obtained by the light curve solution and $q_{SD}$, which uses the assumption of a semi-detached status.
      
   In this study we compared the values of $q_{MS}$, $q_{LC}$ and $q_{SD}$ with the $q$-values as determined by Pourbaix et al. (\cite{Pourbaix1},~http://sb9.astro.ulb.ac.be/). We found the $q_{SD}$-values not representative. The Pourbaix-values are underestimated by $q_{LC}$ and overestimated by $q_{MS}$. The observed mass-ratio distribution of the 303 Algols cited above has hence been recalculated using the mixing of $q_{MS}$ and $q_{LC}$ that represents the numbers in the Pourbaix catalogue best, which leaves still 45$\%$ of the observed Algols in  $q$ $\in$ [0.4-1], as can be seen in Figure \ref{fig_fig2}.
   
\section{Need for liberal evolution}
\label{sec_Liberal}

\begin{figure*}[!ht]
\centering
\includegraphics[width=9.6cm]{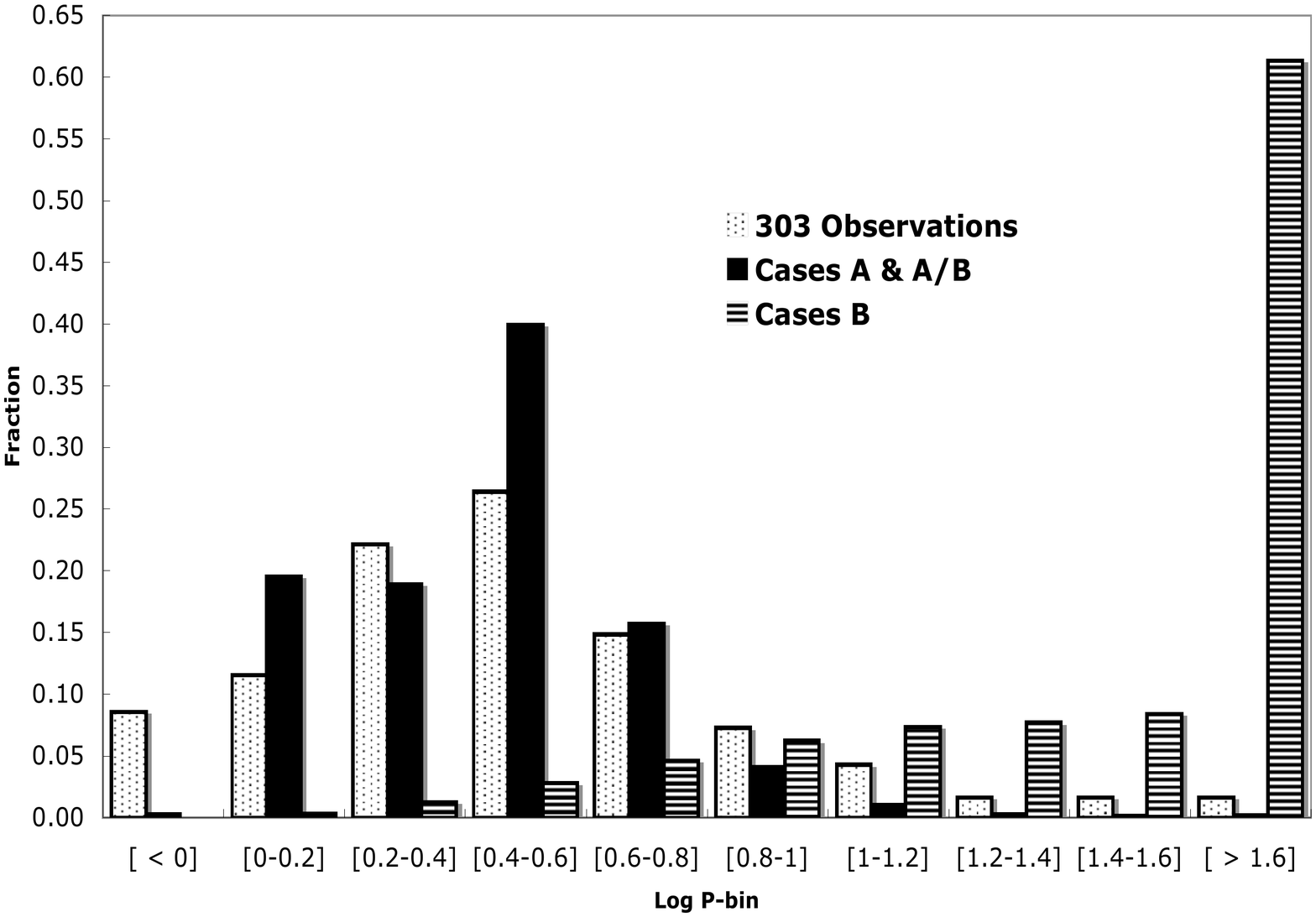}
\caption{Observed distribution of periods of Algols compared to conservative evolution of binaries with a B-type primary at birth.} 
\label{fig_fig1}
\end{figure*}

The $conservative$ simulation has been explained in detail by Van Rensbergen  et al. (\cite{Walter1}). Figure \ref{fig_fig1} compares the observed orbital periods of 303 Algols with a B-type primary at birth with the orbital periods obtained from $conservative$ binary evolution. Larger initial periods leading to case-B RLOF produce mainly Algols with long orbital periods. Cases-A follow the observed distribution better. Since among the Algol population there are far more systems undergoing RLOF-A (fraction of the nuclear time-scale) than RLOF-B systems (fraction of the much shorter Kelvin-Helmholtz time-scale) the observed distribution of orbital periods meets the results from $conservative$ binary evolution well.
   
\begin{figure*}[!ht]
\centering
\includegraphics[width=9.6cm]{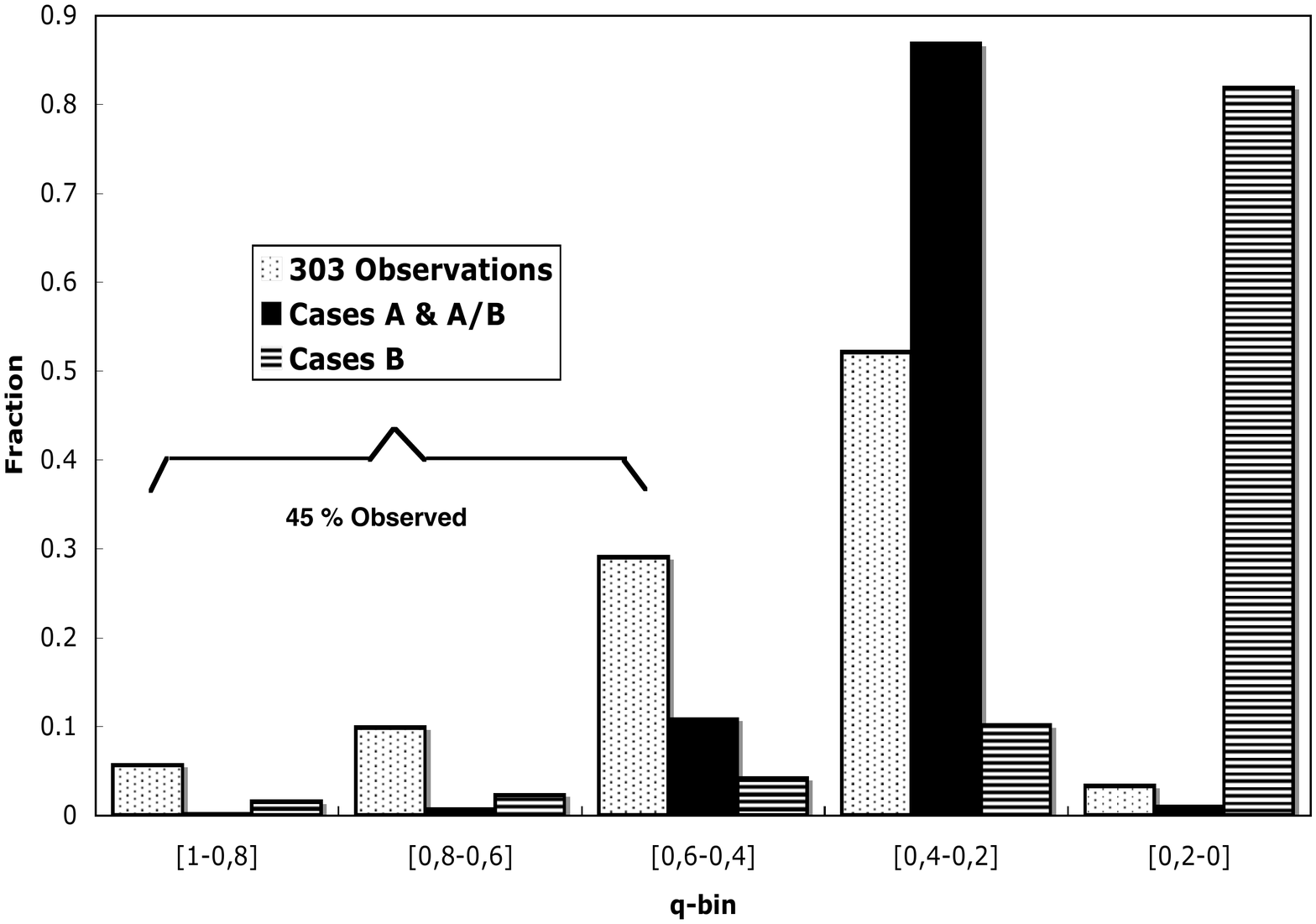}
\caption{Observed distribution of mass-ratios of Algols compared to conservative evolution of binaries with a B-type primary at birth. The fact that $\approx$ 45 $\%$ Algol binaries are observed in $q \in$ [0.4-1] shows that $conservative$ evolution can not always be valid.} 
\label{fig_fig2}
\end{figure*}

Figure \ref{fig_fig2} compares the observed mass-ratios of 303 Algols with a B-type primary at birth with mass-ratios from $conservative$ binary evolution. Cases-B produce more than 80$\%$ Algols with $q$ $\in$ [0-0.2]. Cases-A produce most of their Algols with $q$ $\in$ [0.2-0.4]. The fact that $\approx$ 45 $\%$ of the observed Algol systems are in $q$ $\in$ [0.4-1] forces us to state that $conservative$ evolution can not be the only channel for the evolution of binaries with a B-type primary at birth.

In our search for a $liberal$ scenario we divide interacting binaries into direct impact systems and systems with an accretion disc around the gainer. During their evolution, binaries change positions in the ($r-q$)-diagram, where $q$ is the mass-ratio as defined by relation (\ref{eqn:qratio}) and $r$ the relative radius of the gainer: i.e. its radius divided by the semi major axis of the system. Using the semi-analytical ballistic calculations of Lubow and Shu (\cite{Lubow}), two curves $\omega_{d}$ and $\omega_{min}$ are drawn onto the ($r-q$)-diagram. If a system is located above $\omega_{d}$ the gas flow coming from the donor will hit the gainer directly. If the relative radius of the gainer is below $\omega_{min}$, the gas stream will feed a classical accretion disc with relative outer radius equal to $2 \times\omega_{d}$. We performed our calculations so that systems between $\omega_{d}$  and $\omega_{min}$ will develop an accretion disc with a relative outer radius growing gradually from  $\omega_{d}$ = $ R_{g}$ to $2 \times\omega_{d}$.

Furthermore, the degree of $liberalism$ will be measured with the quantity $\beta$ $\in$ [1-0] which is defined as follows:

\begin{equation}
\dot M_{g}~=-~\beta ~\dot M_{d}^{RLOF} 
\label{eqn:beta}
\end{equation}

$\dot M_{d}^{RLOF}$ is the negative value of the mass lost by the donor and $\dot M_{g}$ the positive value gained by the gainer. $(1-\beta)~\dot M_{d}^{RLOF}$ is hence the negative value $blown~away$ by the gainer. $\beta$~=~1 characterizes $conservative$ evolution, where every amount of mass lost by the donor is captured by the gainer. Figure \ref{fig_fig6} shows an evolution which is $conservative$ most of the time, but in which up to $\approx$ 80 $\%$ (i.e. $\beta$~$\approx$~0.2) of the mass lost by the gainer through RLOF is blown into interstellar space during a $liberal$ era.

\section{Liberal evolution during direct impact}
\label{sec_direct}
\subsection{The rapidly rotating gainer}
\label{sec_rotation}

Conservation of angular momentum spins the gainer up due to the impact of RLOF-material coming from the donor star. Mass located near the gainer's equator gets loosely bound when the gainer rotates rapidly. The  spinning-up of the gainer is characterized by an enhancement of its rotational angular momentum $\Delta J_{g}^{+}$ which is given in cgs-units by Packet (\cite{Packet}), corrected with the impact-parameter ${d}$ as shown in figure \ref{fig_fig4}, simulating the direct-hit scenario (Langer, \cite{Langer}):

 \begin{equation}
     \Delta J_{g}^{+}=6.05\times {10^{51}}\times {\lbrack {R_{g}\over R_{\odot}}\rbrack ^{1 \over 2}}\times {\lbrack{M_{g}\over M_{\odot}}+{{{\Delta M_{g}}\over 2~M_{\odot}} }\rbrack^{1 \over 2}}\times {{\Delta M_{g}}\over M_{\odot}}\times {d\over R_{g}}
\label{upspin}
   \end{equation}

This spin-up is, however, counteracted by tidal interactions which were first studied by Darwin (\cite{Darwin}). The formalism for tidal downspinning can be taken from Zahn (\cite{Zahn}), who gives a suitable approximation for the synchronisation time-scale: 

 \begin{equation}
 \tau_{sync}~(yr)={q^{-2}} \times {\lbrack {a\over R_{g}} \rbrack} ^{6}
\label{tausync} 
  \end{equation}
  
 This expression uses the semi major axis $a$ of the binary and a mass-ratio $q$, in which the star that has to be synchronized is in the denominator. This is the $gainer$ in our case, so that ${q}={M_{d}\over M_{g}}$.

Tidal interactions modulate the angular velocity of the gainer $\omega_{g}$ with the angular velocity $\omega_{orb}$ of the system. According to Tassoul (\cite{Tassoul}) one can write:

 \begin{equation}
{1 \over {\omega_{g}-\omega_{orb}}} \times {d \omega_{orb} \over  d t}= - {1\over {\tau_{sync} \times f_{sync}}}= -{1 \over t_{sync}}
\label{fsyncsync} 
  \end{equation}
  
Using the moment of inertia $I_{g}$ of the gainer we find the expression which was used by Detmers et al. (\cite{Detmers}) in his scenario for the $liberal$ evolution of a massive close binary: 
  
\begin{equation}
     \Delta J_{g}^{-}={I_{g}} \times {(\omega_{orb}-\omega_{g})} \times {\lbrack 1 - e^{({\Delta t  \over  {\tau_{sync} \times f_{sync}}})} \rbrack}
\label{downspin}
   \end{equation}  
  
Tidal interactions spin the gainer down when $\omega_{g} > \omega_{orb}$. Tides spin the gainer up when  $\omega_{g} < \omega_{orb}$. $f_{sync}$ = 1 represents weak tidal interactions whereas $f_{sync}$ = 0.1 implies strong tides.

When the upspinning stops at the end of RLOF, tidal interactions will settle the system into a situation with $\omega_{g} = \omega_{orb}$. Expression (\ref{downspin}) then implies that $\Delta J_{g}^{-}=\Delta J_{g}^{+}=0$. Synchronisation is achieved and angular momentum remains conserved.

\begin{figure*}[!ht]
\centering
\includegraphics[width=9.6cm]{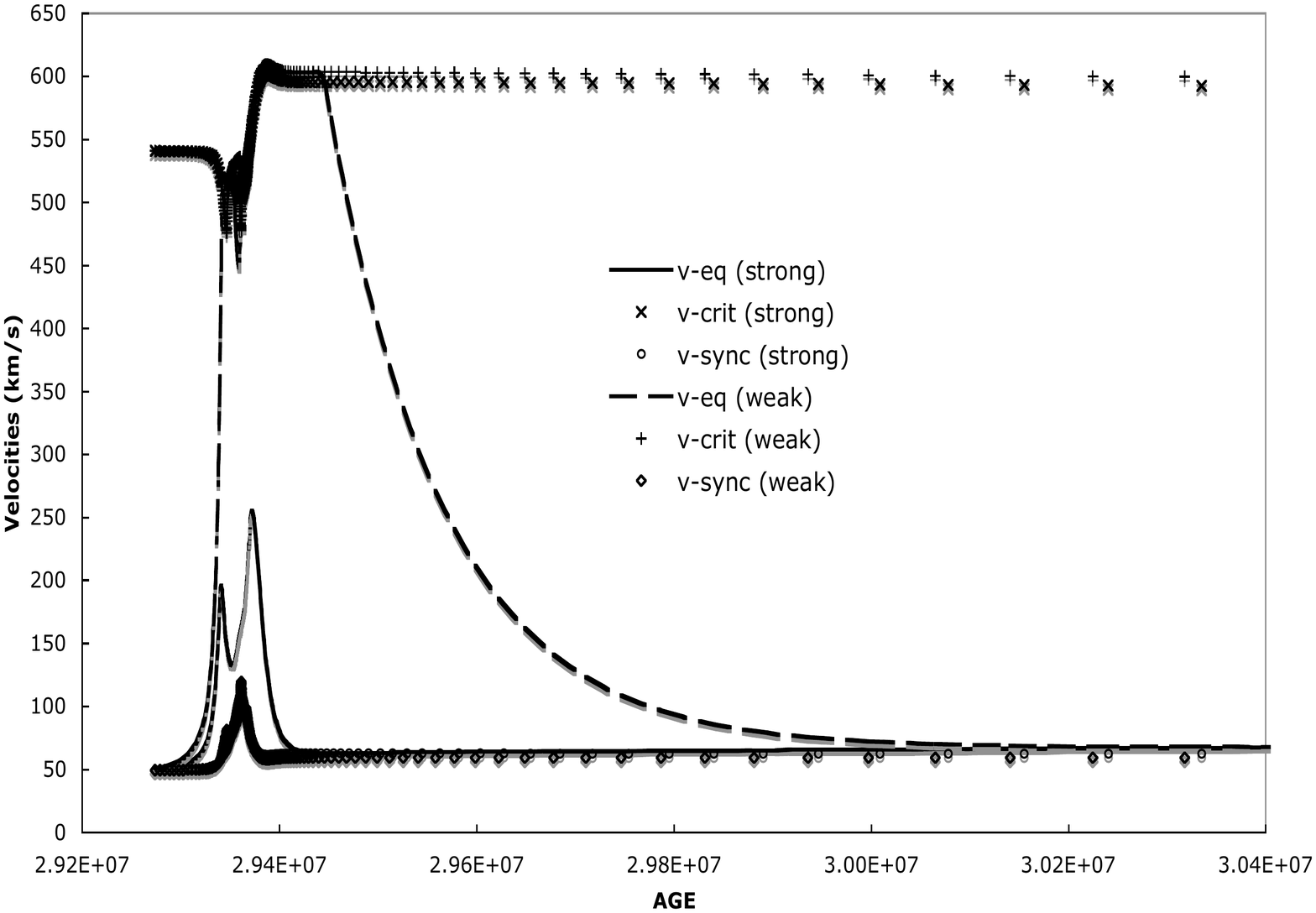}
\caption{The gainer in a (9+5.4) $M_{\odot}$ binary with an initial period of 3.6 days rotates at critical velocity during the rapid phase of mass-transfer if the tidal friction is weak. Strong tidal interaction prevents $very~fast$ rotation of the gainer.} 
\label{fig_fig3}
\end{figure*}

 {The spin-up of the gainer is strongly at work during the short era of rapid mass-transfer soon after the onset of RLOF. Disregarding tidal interactions, Packet (\cite{Packet}) showed that critical velocity at the gainer's equator is achieved more easily if the mass of the gainer is small. Figure \ref{fig_fig3} shows the evolution with time during the rapid phase of mass-transfer of a (9+5.4) $M_{\odot}$ binary with an initial orbital period of 3.6 d. The spin-up works in this case during hydrogen core burning of the donor. From Figure \ref{fig_fig3} it is clear that critical velocity is only achieved assuming weak tidal interactions, whereas strong tides prevent the gainer to rotate at critical velocity.}

\subsection{The hot spot}
\label{sec_hotspot}  

\subsubsection{Visibility of a hot spot}
\label{sec_visible}
  
The accretion luminosity $L_{acc}$ on the gainer's surface is caused by the impact of RLOF-material. This impact causes a bright spot at the trailing side of the gainer's equator (Peters \& Polidan, \cite{Peters}). Such a hot spot can also be created at the outer edge of an accretion disc, leaving the possibility for material to spiral inward. The geometry of the direct impact system shown in figure  \ref{fig_fig4} illustrates that the hot spot is only visible to the observer near phase $\Phi$ $\approx$ 0.75. This is the case as well if the hot spot is located at the gainer's equator or at the edge of an accretion disc.

\subsubsection{Characteristics of a hot spot}
\label{sec_energy}

When matter falls from infinity into the potential well of the gainer it gets hot on the gainer's surface and emits radiation. The accretion luminosity is then given by the difference of the potential energy at infinity and at the point of impact P.

\begin{equation}
{{L_{acc} ^{\infty}= U({\infty}) - U(P) = G\times{M_{g}\times{\dot M}_{d}^{RLOF}\over R_{g}}}}
\label{accluminf} 
\end{equation}	

This accretion luminosity is weakened because matter hits the gainer coming from the first Lagrangian point $L_{1}$ which is different from the idea that it starts from infinity. Taking into account the sonic speed in $L_{1}$, a reduction factor $D \in \rbrack 0-1  \lbrack$ can easily be calculated using the appropriate distances in the corotating system shown in Figure \ref{fig_fig4}. The real accretion luminosity can then be written as $L_{acc} = D \times L_{acc}^{\infty}$ and can be calculated numerically as:

 \begin{equation}
{{L_{acc} \over L_{\odot}} = U(L_{1}) - U(P) = {3.14 \times 10^{7}} \times D \times {{{M_{g}\over M_{\odot}} \times {{\dot M}_{d}^{RLOF} \over M_{\odot} / y}}\over {R_{g} \over R_{\odot}}}} 
\label{acclum} 
\end{equation}

The luminosity of a star is a global quantity. The accretion luminosity $L_{acc}$, however, has to be evaluated locally. If we want matter to escape from the gainer's equator, it has to be removed from the restricted surface area of the accretion zone, which is smaller than the entire gainer's surface. Hence $L_{acc}$ has to be evaluated as a local quantity which is strengthened by the limited area of the accretion zone in which the accretion luminosity is concentrated: $L_{acc} \over S$ can be used as the outward radiative pressure in the area of impact. $S$ is the fraction of the stellar surface covered by the accretion zone ($S$ $\ll$ 1).  $\\$
On the other hand, $L_{acc}$ is weakened by the low efficiency of the accretion luminosity and only  $L_{acc} \over A$ ($A$ $\gg$ 1) can be converted into radiative flux.

The radiation pressure exercised by a hot spot is thus produced by :

 \begin{equation}
{L_{acc} \over {S \times A}} = {{L_{acc} \over {K}} {~~with~~ K = S \times A}}
\label{Kfactor} 
\end{equation}

The numbers $A$ and $S$ (and hence the crucial quantity $K$) can be derived from the observations using the following procedure.  $\\$
When the gainer has no spots its luminosity is $L_{g}^{0}$. The luminosity of the spotted gainer is then given by:

 \begin{equation}
{{L_{g}^{1}\over L_{\odot}} = {{L_{g}^{0}\over L_{\odot}} + {L_{add}\over L_{\odot}}}} 
\label{Spotlum} 
\end{equation}

$L_{add}$ is the fraction of the accretion luminosity $L_{acc}$, given by relation ($\ref{acclum}$), which is converted into radiative flux. 
\begin{flushleft}
$\bullet$~~~$L_{add}$ $<$ 0 implies a dark spot\\ 
$\bullet$~~~$L_{add}$ = 0 implies a non spotted stellar surface\\
$\bullet$~~~$L_{add}$ $>$ 0 implies a hot spot. \\  
\end{flushleft}

We assume that the hot spot is created by RLOF-material infalling from the donor star.
 
Next, we can rewrite equation (\ref{Spotlum}) as: 

 \begin{equation}
{{L_{g}^{1}\over L_{\odot}} = \lbrack {{S_{g}-S_{spot}} \over S_{\odot}} \times {{T_{eff,g}^{4}} \over (5770)^{4}} \rbrack~+ \lbrack {{S_{spot}} \over S_{\odot}} \times {{T_{spot}^{4}} \over (5770)^{4}} \rbrack} 
\label{Spotlumnew} 
\end{equation}

For a hot spot with $T_{spot}$ $>$ $T_{eff,g}$ we can write this as:

 \begin{equation}
{L_{add}\over L_{\odot}} = {{S_{spot}\over S_{\odot}} \times {1 \over (5770)^{4}} \times \lbrack T_{spot}^{4} - T_{eff,g}^{4} \rbrack}
\label{Addlum} 
\end{equation}

Knowledge of $L_{add}$ (e.g. through direct observation) determines the quantity $A$ as: ${L_{add} = {L_{acc} \over A}}$. Since $L_{add}$ $\ll$ $L_{acc}$, we have that A $\gg$ 1.

Temperature and size of the spot are related through relation (\ref{Addlum}), which can be transformed into:

\begin{equation}
{T_{spot}^{4}=T_{eff,g}^{4}+ {\lbrack {L_{add} \over L_{\odot}}} \times {S_{\odot} \over S_{spot}} \times {(5770)^{4}} \rbrack}
\label{Tspot} 
\end{equation}

Every measured value $L_{acc}$ is reproduced by an infinite number of combinations of $S_{spot}$ and $T_{spot}$. Only a restricted range of values hereby reproduce a realistic hot spot. Knowledge of $S_{spot}$ (e.g. through direct observation) determines the quantity $S$ as $S_{spot} \over S_{g}$. Since $S_{spot}$ $\ll$ $S_{g}$, we have that S $\ll$ 1. The crucial quantity $K$, as defined by equation (\ref{Kfactor}), is thus the product of a factor $A$ $\gg$ 1 and a factor $S$ $\ll$ 1. It is clear that small values of $K$ will support mass-loss whereas large values of $K$ will suppress mass-loss from the system.

It has to be noticed that the quantity $K$ can be evaluated directly from equation (\ref{Addlum}) which can be written as:

 \begin{equation}
{{L_{acc}\over {K  \times L_{\odot}}} = {{\lbrack {R_{g} \over R_{\odot}}\rbrack}^{2}} \times {{\lbrack {1 \over 5770}\rbrack}^{4}} \times {\lbrack T_{spot}^{4} - T_{eff,g}^{4}} \rbrack }
\label{Knew} 
\end{equation}

One can thus consider $K$ as a function of $L_{acc}$ and $T_{spot}$, a relation that determines the important quantity $K$ without disentangling it into its constituents $A$ and $S$. Measurement of the temperature of the spot and determination of the mass-transfer-rate (often using observed changes of the orbital period), which determines the accretion luminosity through relation (\ref{acclum}) enable us to determine the quantity $K$ throughout the entire evolutionary computation.

\subsection{Mass can leave the system}
\label{sec_splash}

If one thinks about possible mass-loss from the gainer into the interstellar medium we first have to evaluate the binding energy of a test mass $m$ in the hot spot point $P$ of Figure  \ref{fig_fig4}.  Without rotation of the gainer and without accretion luminosity the energy of a test mass $m$ at rest in $P$ is given by:

\begin{figure*}[!ht]
\centering
\includegraphics[width=9.6cm]{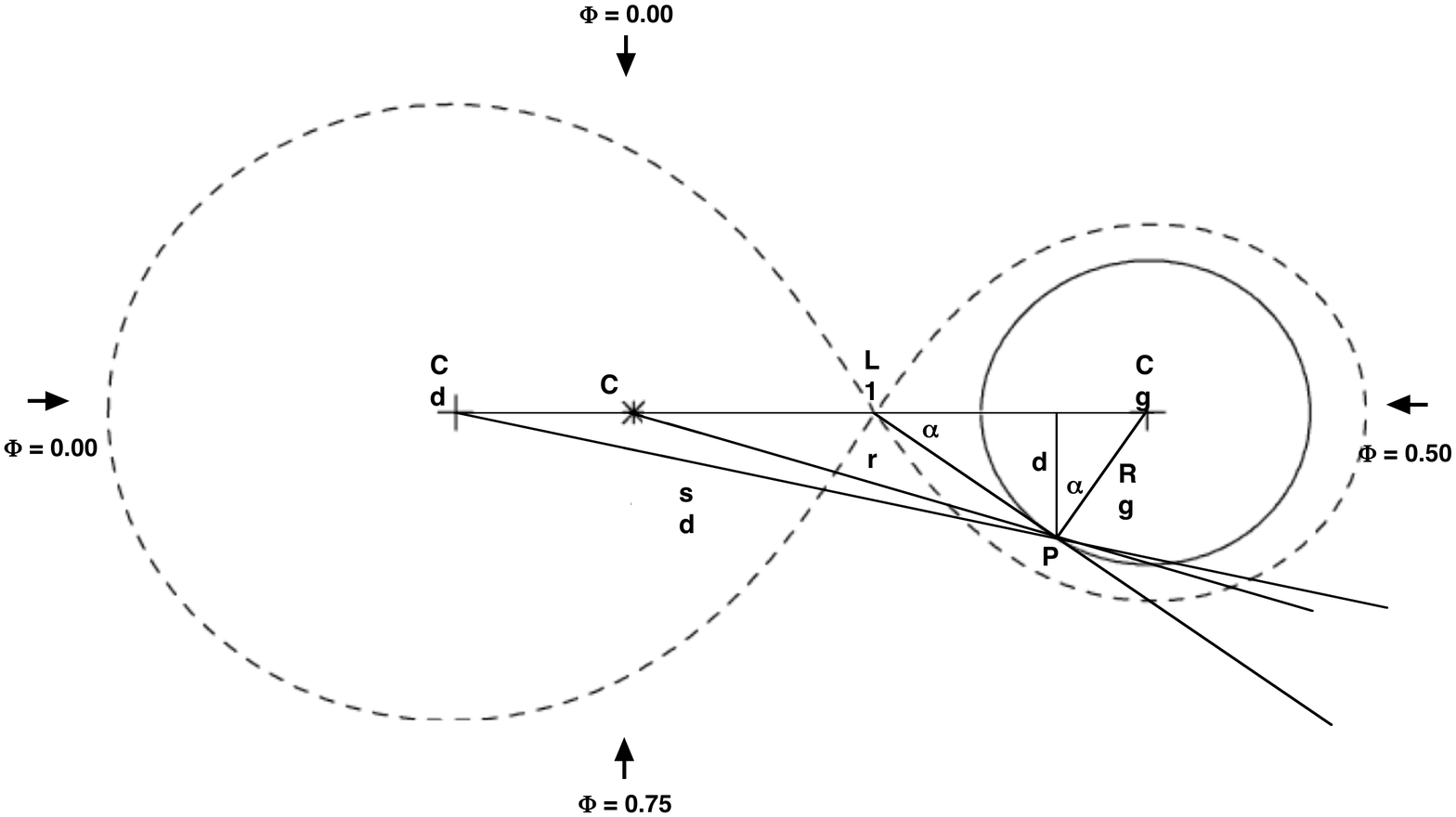}
\caption{Geometry determining the binding energy of a test mass in the impact point $P$ in the middle of a hot spot. The observer looks into the direction of the small arrows. The hot spot $P$ is turned towards the observer only near phase $\Phi$ $\approx$ 0.75.} 
\label{fig_fig4}
\end{figure*}

 \begin{equation}
{{U}= {-~G ~\lbrack {{M_{d} ~ m} \over {s_{d}}}} + {{M_{g} ~ m} \over {R_{g}}}\rbrack~-{1\over 2}~m~{r^{2}}~{\lbrack {{2~\pi}\over {P_{orb}}}\rbrack^{2}}}
\label{BindingU} 
\end{equation}

This negative amount of energy is counteracted by the luminosity of the gainer ($L_{nucl}$), its enhanced equatorial velocity and the positive energy input $L_{acc}$ given by relation (\ref{acclum}). The impacting gas stream has, however, already been used to spin up the gainer as explained in section (\ref{sec_rotation}). The rotational kinetic energy is hereby raised at a rate $\dot K_{rot}$. The energy rate which remains available to build the hot spot is thus a little less than $L_{acc}$ and equal to:

 \begin{equation}
{{{L_{acc}^{-}}} = {L_{acc}-{\dot K_{rot}}}} 
\label{lummin} 
\end{equation}	

This remaining part ($L_{acc}^{-}$) is used to increase the temperature of the hot spot. When the rotational energy ${{1\over 2}~m~v_{eq}^{2}}$ of a test mass located at the gainer's equator approaches the break-up velocity already almost 50 $\%$ of the binding energy given by relation (\ref{BindingU}) is surmounted.

The quantity $\dot K_{rot}$ is not easy to observe and has thus been neglected in section (\ref{sec_energy}) but is easily followed up during a run of the  {\it Brussels binary evolution code}. When the gainer rotates rapidly, the spinning up becomes almost impossible and practically all the energy available from the accretion will then be used for building the hot spot, so that ${L_{acc}^{-} \approx L_{acc}}$.

The work done by the outward radiative force is:

 \begin{equation}
 K_{rad}= {{m~{\bar \kappa}}\over c}~{\lbrack {{L_{nucl} + {(L_{acc}^{-})\over {K}}} \over {4~\pi~R_{g}}}}\rbrack 
 \label{kappa} 
\end{equation}	
 
One obtains the total amount of energy on a test particle at the gainer's equator by adding the rotational kinetic energy and the energy input as given by relation  (\ref{kappa}) as positive terms to equation (\ref{BindingU}). The test mass can only leave the system, when this energy is positive. So matter in the hot spot can only leave the system if (after convenient division by $m$):

\begin{equation}
{{-G \lbrack {{M_{d}} \over {s_{d}}}} + {{M_{g}} \over {R_{g}}}\rbrack-{1\over 2}~{r^{2}}{\lbrack {{2\pi}\over {P_{orb}}}\rbrack^{2}}+{{{1\over 2}v_{eq}^{2}}+{{{\bar \kappa}~{({L_{acc}^{-}\over K}}+L_{nucl})}\over{4~\pi~c~R_{g}}} > 0}}
\label{Finalcondition} 
\end{equation}

$\dot M_{d}^{RLOF,crit}$ is that value of $\dot M_{d}^{RLOF}$ which causes a value of $L_{acc,crit}$ such that the left hand side of equation (\ref{Finalcondition}) equals zero. Every amount of $\dot M_{d}^{RLOF}$ in absolute value exceeding $\dot M_{d}^{RLOF,crit}$ will leave the system. In other words: {\it supercritical mass will be lost by the system}. 

For simplicity one could take only the Thomson scattering into account to evaluate relation  (\ref{Finalcondition}):

\begin{equation}
\bar \kappa =  \kappa_{e} = 0.34~cm^{2} g^{-1}
\label{Thomson} 
\end{equation}

In this paper we have replaced $\bar \kappa$ by $\kappa_{Ross}$. The Rosseland opacities at the gainer's surface for solar abundances and the appropriate values of log $T_{eff}$ and log g were hereby taken from Kurucz (\cite{Kurucz}), defining the quantity F as:

\begin{equation}
F~=~{\kappa_{Ross} \over \bar \kappa_{e}} 
\label{Ross} 
\end{equation}

\begin{figure*}[!ht]
\centering
\includegraphics[width=9.6cm]{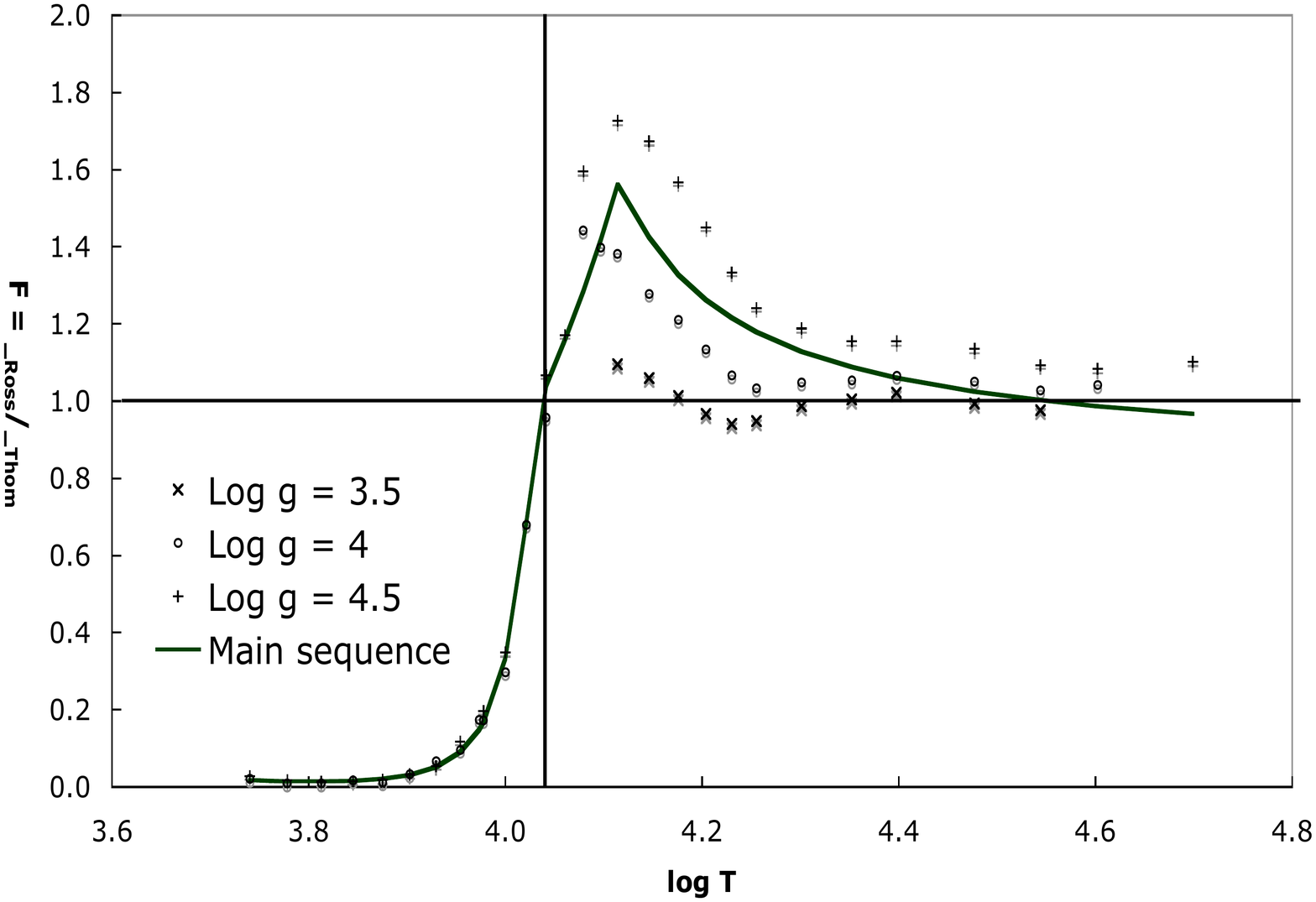}
\caption{Opacity ratio ${\kappa_{Ross} \over \bar \kappa_{e}}$ as a function of temperature, showing the dominance of $\kappa_{Ross}$ in a temperature region above 11,000 K.} 
\label{fig_fig5}
\end{figure*}

Figure \ref{fig_fig5} shows that at low temperatures the Rosseland opacity is smaller than given by relation (\ref{Thomson}), whereas Thomson scattering produces the total opacity at high temperatures. From 11,000 K on, however, there is a temperature range wherein the Rosseland opacity is the largest, so that it demands less energy for the  test particle in the hot spot to leave the system. Since the temperature of the hot spot exceeds the temperature of its surroundings, we calculate in our $Brussels~binary~evolutionary~code$ the quantity $F$ at a temperature equal to $1.5 \times T_{eff,g}$.

With these assumptions, a  numerical value for $\dot M_{d}^{RLOF,crit}$ can be found from equations (\ref{Finalcondition}) and (\ref{acclum}):

\begin{eqnarray}
{\vert \dot M_{d}^{RLOF,crit}\vert} &=& {{K ~ R_{g}^{2}} \over {F ~ D ~ M_{g}}} ~ \lbrace 1.224~10^{-3}{\lbrack{M_{d} \over s_{d}}+{M_{g} \over R_{g}}\rbrack} \nonumber \\
&+&  8.219~10^{-6}~\lbrack {r \over P_{orb}} \rbrack ^{2} - {3.208~10^{-9}~v_{eq}^{2}} \nonumber \\
&+& {3.183~10^{-8}~{F\over K~R_{g}}~\dot K_{rot}}  \nonumber \\
&-& {3.183~10^{-8}~{F\over R_{g}}~L_{nucl}} \rbrace
\label{FinalMdot} 
\end{eqnarray}

Every quantity is expressed in the usual units and $v_{eq}$ in ${km\over s}$.

\vspace{0.2cm}

Mass-loss out of a binary is mainly driven by rapid rotation and large mass-transfer-rates for systems with a B-type primary at birth. For very massive binaries, however, $L_{nucl}$ may be sufficiently large so as to drive matter out of a binary in which the gainer is spinning rapidly (Petrovic et al. \cite{Petrovic}).

\section{Possible liberal evolution with an accretion disc}
\label{sec_disc}

 {The geometry of a binary with an accretion disc was introduced in section (\ref{sec_Liberal}), yielding accretion discs with outer radius $R_{disc}$ between $R_{g} = \omega_{d} \times a $ and $2 \times \omega_{d} \times a$. Replacing $R_{g}$ with $R_{disc}$ relations (\ref{accluminf}) and (\ref{acclum}) can be used to calculate the accretion luminosity in the point $P$ at the outer edge of the spot. This accretion luminosity is considerably reduced when the distance between $L_{1}$ and $P$ is small.}

 {For the calculation of the critical amount of mass-transfer, forcing the system to evolve in a $liberal$ way, equation (\ref{FinalMdot}) has to be modified. Dealing with a Keplerian accretion disc, the equatorial velocity of the gainer has to be replaced by the Keplerian velocity of a test mass located at the edge of the disc. Due to the virial theorem, the rotational kinetic energy of the test mass covers exactly 50 $\%$ of the gravitational potential energy caused by the gainer. In this way, one can approximate the quantity $\dot M_{d}^{RLOF,crit}$ with:}

\begin{eqnarray}
{\vert \dot M_{d}^{RLOF,crit}\vert} &=& {{K ~ R_{disc}^{2}} \over {F ~ D ~ M_{g}}} ~ \lbrace 1.224~10^{-3}{\lbrack{M_{d} \over s_{d}}+{M_{g} \over 2 \times R_{disc}}\rbrack} \nonumber \\
&+&  {8.219~10^{-6}~\lbrack {r \over P_{orb}} \rbrack ^{2}}  \rbrace
\label{Mdotdisc} 
\end{eqnarray}

 {Our calculations as summarized in section (\ref{sec_Results}) show that initially (9+5.4) $M_{\odot}$ binaries evolve almost always $conservatively$, except during a short era of rapid mass-transfer short after the onset of RLOF. This situation occurs always during hydrogen core burning of the $donor$ and with the $direct~impact$ geometry. The accretion disc is only formed when the orbit widens. The mass-transfer-rate is, however, not sufficiently large in this case so that the value given by expression (\ref {Mdotdisc}) is never attained. With our simple assumptions, disc systems that are formed after an era of RLOF-A, evolve $conservatively$. Intermediate-mass binaries with larger initial orbital periods so that RLOF-A is avoided and an accretion disc is formed (not included in this paper) might very well also experience an era of $liberal$ evolution.}

 {The mass-loss rates in Table 1 cover the direct impact systems only.  Mass-loss in the case of an accretion disc around the gainer would enhance the degree of $liberalism$ mentioned in Table 1. Subsection (\ref{sec_accretion disc}) mentions, however, several intermediate-mass binaries with an accretion disc by which mass-loss from the system was observed. Hydrodynamical calculations by Bisikalo et al. (\cite{Bisikalo}) follow matter that spirals inwards and creates a subsequent hot line, caused by the shock issued from the interaction of the circumstellar disc with the flow from the inner Lagrange point $L_{1}$. This concept was introduced in three-dimensional numerical simulations by Sytov et al. (\cite{Sytov}) for a system with an orbital period of 0.23 d, a low-mass donor of 0.56 $M_{\odot}$ and a 0.6 $M_{\odot}$ white dwarf gainer. Even this low-mass system with an accretion disc around the gainer loses mass at a rate of  $\approx$ 3 $\times$ $10^{-10}$ $M_{\odot} \over y$ through the third Lagrangian point $L_{3}$. This type of mass-loss is $not~yet$ included in our computational code.}

\section{Calibrating the quantities $A$ and $S$}

In this section we will evaluate relation (\ref{FinalMdot}) carefully.

\begin{flushleft}
$\bullet$~~$D={L_{acc} \over L_{acc}^{\infty}}$ is followed up during the entire evolutionary calculation\\ 
\vspace{0.2cm}
$\bullet$~~$F={ \kappa_{Ross}  \over \bar \kappa_{e}}$ is given introducing $T_{spot}$ into figure \ref {fig_fig5}\\ 
\end{flushleft}

The quantity $K$ is, however, more difficult to determine. For only a few binaries observations of $L_{acc}$, $L_{add}$, $S_{spot}$  are available in order to determine the quantities $A$ and $S$ and subsequently $K$:

\begin{flushleft}
$\bullet$~~~$A={L_{acc} \over L_{add}}$\\ 
\vspace{0.2cm}
$\bullet$~~~$S={S_{spot} \over S_{g}}$\\
\vspace{0.2cm}
$\bullet$~~~$K = A \times S$ \\  
\end{flushleft}

The situation where $T_{spot}$ has been observationally determined is largely to be preferred, because the quantity $K$ can then be directly evaluated using relation (\ref{Knew}), which can also be written as:  

\begin{equation}
K ={{{L_{acc} \over L_{\odot}} \times {(5770)^{4}}} \over {{{\lbrack {R_{g} \over R_{\odot}}\rbrack}}^{2}} \times  {\lbrack  {{T_{spot}^{4}} - {{T_{eff,g}}^{4}} \rbrack}}}
\label{Knewest} 
\end{equation}

It is clear that the value of $K$ using relation (\ref{Knewest}) can differ from the ($K=A\times S$)-calculation because all observed stellar parameters are not always entirely internally consistent.

\subsection{Gauging the quantity $S$}

Gunn at al. (\cite{Gunn}) use an expression of Pringle (\cite{Pringle}) giving the area of the cross section of the stream starting from the first Lagrangian point $L_{1}$ with the local sonic speed $v_{sonic}$ towards the gainer, for the calculation of the surface area of a hot spot around the impact point P:

\begin{equation}
{S_{spot}~=~2.77~{10^{18}}~{T_{eff,d}^{1 \over 2}}~{\vert L_{1}P \vert \over R_{\odot}}~{\lbrack {R_{d}\over R_{\odot}}\rbrack^{3\over 2}}~{\lbrack {M_{d}\over M_{\odot}}\rbrack^{-{1\over 2}}}~~cm^{2}}
\label{Sspot} 
\end{equation}

In this expression (\ref{Sspot}) we have used $v_{sonic}~=~0.125~\sqrt {T}~{km\over s}$. The number $0.125$ has been taken between $0.1$ and $0.15$, numbers which are valid using solar abundances for respectively a neutral gas and a completely ionized plasma. The quantity $S={S_{spot} \over S_{g}}$ can now be calculated for every binary. Surface areas of the spot have been determined for a few binaries:  VW Cep (Pustylnik \& Niarchos \cite{Pustylnik}), CN And (Van Hamme et al. \cite{Vanhamme}), KZ Pav (Budding et al. \cite{Budding})  and V505 Sgr (Gunn et al. \cite{Gunn}), showing a fair agreement with the results from expression (\ref{Sspot}). 

\subsection{Gauging the quantity $A$}
\label{sec_quantA}

Section (\ref{sec_quant_K}) lists the semi-detached binaries for which observations are available allowing to determine some or all of the quantities $A$, $S$ and $K$. The cases for which $S (S_{spot}),~A (L_{add})~and~K (T_{spot})$ were measured or determined separately but for which the relation $K = A\times S$ was obviously violated are not included.

The quantity $A$ is defined from the measured quantity $L_{add}$ and the calculated quantity $L_{acc}$, as $A={L_{acc} \over L_{add}}$. $L_{acc}$ is determined in our sample of semi-detached binaries in section (\ref{sec_quant_K}) using the values of $dP \over dt$ which were determined with the widely used (O-C)- procedure explained in detail by Sterken (\cite{Sterken}). Observed times of light minima (O) of eclipsing binaries were taken from the website of Kundera (\cite{Kundera}, \verb"http://www.oa.uj.edu.pl/ktt/krttk_dn.html"), completed for the years after 2002, with the data published in various issues of the "Information Bulletin on Variable Stars, Commission 27 of the I.A.U.", which we found for each binary at \verb"http://simbad.u-strasbg.fr/simbad/". Subsequently, a best fit parabola was drawn through the observed times of light minima (C). The (O-C)- procedure then yields the $observed$ value of  $dP \over dt$ (expressed in in ${d \over y}$). The parabolic fit is sometimes convincing and sometimes doubtful. The latter cases have been disregarded because they occur when mechanisms different from mass-transfer (e.g. magnetic braking) are at work.

When a convincing parabolic behavior $dP \over dt$ is available, the negative value of $dM_{d} \over dt$ (expressed in in ${{M_{\odot}} \over y}$) was calculated with the $conservative$ relation (\ref{dM/dt}). We use this value of  $dM_{d} \over dt$ to calculate $L_{acc}$ with relation (\ref{acclum}). 

\begin{equation}
{-~{dP \over P}=3{dM_{d} \over M_{d}}+3{dM_{g} \over M_{g}}}
\label{dM/dt} 
\end{equation}

This equation is the conservative limit of the more general relation including mass-loss  (dM $\neq$ 0) and angular momentum loss  (dJ $\neq$ 0) from the system.

\begin{equation}
{-~{dP \over P}=3{dM_{d} \over M_{d}}+3{dM_{g} \over M_{g}}-3{dJ \over J}-{dM \over M}}
\label{dM/M} 
\end{equation}

The conservative relation (\ref{dM/dt}) has been used in section (\ref{sec_quant_K}) because most of these systems are in a quiet phase of conservative mass-transfer. The strict application of the conservative limit in liberal cases underestimates the real amount of mass-transfer only very slightly (Erdem et al. \cite{Erdem}), the difference being well below the uncertainty introduced by the (O-C)-procedure.

\section{Calibration of the quantity $K$}
\label{sec_quant_K}

In this section we list the few semi-detached binaries for which one or more of the quantities $A,~S~and~K$ have been determined within reasonable limits. Since $K= A\times S$, two of the three quantities allow to determine the third one. Simultaneous measurement of the three quantities is, however, very useful in order to narrow the error bars around the individual estimates. We remind the reader that the quantity $A$ is known from $L_{add}$, the quantity $S$ from $S_{spot}$ and the quantity $K$ from $T_{spot}$. It is obvious that more accurate and a greater number of observations would highly improve the statistics outlined below. Especially accurate determinations of spot temperatures would make a more precise determination of the crucial quantity $K$ possible. We have sorted the systems by ascending values of their total mass.

\subsection{Direct impact systems}

\subsubsection{VW Cep}

Stellar parameters and restricted ranges for $L_{add}$, $S_{spot}$ and $T_{spot}$ from Pustylnik \& Niarchos (\cite{Pustylnik}), internally consistent values of ${dP \over dt}$ and ${dM \over dt}$ from Devita et al. (\cite{Devita}) and Pribulla et al. (\cite{Pribulla}) determine the value of K in the range [0.1-0.2].

\subsubsection{CN And}

Stellar parameters from Van Hamme et al. (\cite{Vanhamme}), ${dP \over dt}$ from Samec et al. (\cite{Samec}), spot temperatures and sizes from Van Hamme et al. (\cite{Vanhamme}) and references therein determine the value of K in the range [0.08-0.18].

\subsubsection{KZ Pav}

Stellar parameters and approximate values of $L_{add}$ and ${dP \over dt}$ and thus ${dM \over dt}$ are given by Budding et al. (\cite{Budding}). Consistent values of ${dP \over dt}$ and ${dM \over dt}$ can be found in Walker \& Budding (\cite{Walker}). The value of $S_{spot}$ can been calculated with the expression  (\ref{Sspot}) or can be calculated using the isomorphy with the very similar systems CN And and VW Cep for which $S_{spot}$ has been measured.  Values of $T_{spot}$ found from $L_{add}$ and $S_{spot}$ are comparable to those of CN And and VW Cep. We find a value of $K$ in the range [0.03-0.08].

\subsubsection{V361 Lyr}

Stellar parameters are given by Hilditch (\cite{Hilditch}) and Yakut \& Eggleton (\cite{Yakut}). The mass-transfer-rate has been established by Hilditch (\cite{Hilditch}) as $2.18 \times 10^{-7}$ $M_{\odot} \over y$, who also gives a value of $T_{spot}$ = 9,500 K and $L_{add} \over L_{\odot}$ = 0.5. These numbers generate $A$ = 2.6 and $K$=0.36. From this one finds $S$=0.14, which is more than one order of magnitude larger than the surface area of the spot given by relation (\ref{Sspot}): $S$ = 0.011. Since relation (\ref{Sspot}) yields surface areas that often overestimate the real spot surface area, a spot surface area covering 14 $\%$ of the whole stellar surface is extremely high. With $S$ = 0.011 we find $K$ = 0.028 and $T_{spot}$ = 17,700 K. Hence we have taken a mean value of $K$ for calibration.

\subsubsection{RT Scl}

Stellar parameters and $T_{spot}$ are given by Banks et al. (\cite{Banks}). A narrow range of possible values of ${dP \over dt}$ and hence ${dM \over dt}$ can be found in Clausen \& Gronbech (\cite{Clausen}), Rafert \& Wilson (\cite{Rafert}) and Duerbeck  \& Karimie (\cite{Duerbeck}). Using these data we find $K$ in the range [0.02-0.04]. With the value of $S$ given by expression (\ref{Sspot}) we also determine the efficiency factor $A$ in the range [3-6]. 

\subsubsection{U Cep}

Stellar parameters are taken from Budding et al. (\cite{Budding}). Pustylnik (\cite{Pustylnik2}) quotes a mass-transfer-rate of $\approx$ $10^{-6}$ $M_{\odot} \over y$ but we will use the slightly smaller value $5 \times 10^{-7}$ $M_{\odot} \over y$ as derived with the (O-C) method as outlined in section (\ref{sec_quantA}). A hot spot is reported from observations with the $FUSE$ spacecraft by Peters  (\cite{PetersA}).  The spot temperature is found to be $\approx$ 30,000 K, leading to $K$ $\approx$ $3.3  \times 10^{-3}$ for this system. The impact on the surface of the gainer creates a splash zone in which large velocities are identified in the spectrum. In order to keep the quantity $A$ within physically possible limits, the impact zone should in this case be an order of magnitude smaller than $S$ $\approx$ 0.02, as reported by Peters (\cite{PetersA}).

\subsubsection{U Sge}

A consistent set of stellar parameters is given by Kempner \& Richards (\cite{Kempner}), Richards \& Albright (\cite{Richards}) and Vesper et al. (\cite{Vesper}). Manzoori \& Gozaliazl (\cite{Manzoori}) have determined a mass-transfer-rate of $1.79  \times 10^{-6}$ $M_{\odot} \over y$. A hot spot with a temperature in the range [20,000 - 100,000] K has been seen by Richards \& Albright (\cite{Richards}). Hot circumstellar gas at temperatures in the range [60,000-200,000] K has been identified by Kempner \& Richards (\cite{Kempner}). With a hot spot surface area given by relation (\ref{Sspot}) only the lowest possible spot temperature of $\approx$ 20,000 K can be accepted. Higher spot temperatures violate the law of conservation of energy, because they generate values of the quantity $A$ below 1. Consequently we find a value of $K$ $\approx$ 0.02.

\subsubsection{SV Cen}

Stellar parameters as given by Brancewicz \& Dworak (\cite{Brancewicz et al.}) predict an orbital period increase rather than the observed decrease. Stellar parameters from Wilson \& Star (\cite{WilsonStar}) and Drechsel et al. (\cite{Drechsel}) yield contact systems in which there is no room for the construction of a hot spot. A small sized hot spot with a high temperature (T $\approx$ $10^{5}$) has, however, been observed by Drechsel et al. (\cite{Drechsel}) in the UV. In this section we have used the stellar parameters of Rucinski et al. (\cite{Rucinski}) which yield a close semi-detached binary. Herczeg \& Drechsel (\cite{Herczeg}) notice that ${dP \over dt}$ is very unsteady giving a value of ${dM \over dt}$ in the range [1-4] $\times$ $10^{-4}$ $M_{\odot} \over y$. But even such high mass-transfer-rates cannot create a temperature of $10^{5}$ K in a hot spot with a size as given by relation (\ref{Sspot}). Depending upon the mass-transfer-rate we obtain spot temperatures in the range [30,000-45,000] K. The largest value of $K$ that can be obtained is this way and will be used in this analysis is $\approx$ 4.64 $\times$ $10^{-3}$. The spot temperature of $\approx$ $10^{5}$ K would originate in a very small splash zone characterized by $K$ $\approx$ $10^{-4}$ which according to relation (\ref{FinalMdot}) would lower the critical amount of mass-transfer below any reasonable value.

\subsection{Systems with an accretion disc}
\label{sec_accretion disc}

In this case the hot spot is created at the edge of an accretion  disc. The surface area of the spot is now hardly defined by relation (\ref{Sspot}). The quantity $K$ can now be derived by comparing the temperature of the hot spot with the temperature of the edge of the disc rather than with the surface of the gainer. The radius of the accretion disc has been calculated as explained in section (\ref{sec_Liberal}). For the temperature profile of the disc we take the formulas as derived from first principles by Carroll \& Ostlie (\cite{Carroll}):

\begin{eqnarray}
{T(r)=T_{disc}~ \lbrack {R_{g} \over r} \rbrack^ {3\over4}\lbrack {1-\sqrt{R_{g} \over r}} \rbrack^ {1\over4}}~ \nonumber \\
{T_{disc}~=~{478,074~{\lbrack {{M_{g} \vert \dot M_{d}^{RLOF} \vert}\over {R_{g}^{3}}}}}\rbrack^{1\over4}}
\label{Tdisc} 
\end{eqnarray}

Masses and radii are in solar units and mass-transfer-rates in solar masses per year. The systems are again sorted by ascending values of their total mass.

\subsubsection{KO Aql}

Stellar parameters are taken from Soydugan et al. (\cite{Soydugan}) and Vesper et al. (\cite{Vesper}).  The data of Panchatsaram \& Abhyankar (\cite{Panchatsaram}) yield a detached system and have therefore been disregarded. ${dP \over dt}$ is taken from Panchatsaram \& Abhyankar (\cite{Panchatsaram}) and Soydugan et al. (\cite{Soydugan}). This system has an accretion disc with $R_{disc} \approx 1.5 \times R_{g}$. Expression (\ref{Tdisc}) calculated with  ${dM \over dt}$, $M_{g}$ and $R_{g}$ yields a disc temperature of $\approx$ 9,500 K which drops to $\approx$ 4,600 K at the edge. Soydugan et al. (\cite{Soydugan}) have observed hot spots in this system in which the gainer spins 30 $\%$ faster than synchronously. The quantity $K$ can unfortunately not be estimated without measurements of $T_{spot}$.

\subsubsection{SW Cyg}

This system with an accretion disc shows much similarity with TT Hya. A consistent set of stellar parameters has been given by Richards \& Albright (\cite{Richards}), Budding et al. (\cite{Budding et al.}), Brancewicz et al. (\cite{Brancewicz et al.}) and Vesper et al. (\cite{Vesper}). Qian et al. (\cite{Qianb}) have determined $dM \over dt$ = $2.15 \times 10^{-7}$ $M_{\odot} \over y$ from a measured value of $dM \over dt$. Relation (\ref{Tdisc}) yields $T_{disc}$ = 6,300 K and $T_{edge}$ = 3,100 K.  Albright \& Richards (\cite{Albright}) found that the outer edge of the disc of TT Hya emits 5 times as much $H_{\alpha}$ radiation than the same structure of SW Cyg. Gauging this with the spot temperature of TT Hya we find for SW Cyg: $T_{spot}$ $\approx$ 11,300 K. The value of $K$ is in this case located around 0.016.

\subsubsection{TT Hya}

A coherent set of stellar parameters for this system is given by Eaton \& Henry (\cite{Eaton}), Peters  \& Polidan (\cite{Peters}), Van Hamme \& Wilson (\cite{VanHamme2}), Richards \& Albright (\cite{Richards}), Kulkarny \& Abhyankar (\cite{Kulkarny}) and Panchatsaram \& Abhyankar (\cite{Panchatsaram}). The geometry of the system is thus well defined and a detailed hydrodynamical model is given by Miller et al. (\cite{Miller}). Peters \& Polidan (\cite{Peters}) quote a small value of $L_{add}$ and a spot temperature of 17000 K. Unfortunately there are no measures available of $dP \over dt$ which makes a reliable determination of $dM \over dt$ difficult. Peters \&  Polidan (\cite{Peters0}) argue a lower limit of $10^{-12}$ $M_{\odot} \over y$, whereas Miller et al. (\cite{Miller}) determine a somewhat larger mass-transfer-rate of $\approx$ $2 \times 10^{-10}$ $M_{\odot} \over y$, which agrees with the low value of $L_{add}$ but contradicts the high spot temperature and the disc temperature which is evaluated at 7,000 K by Miller et al. (\cite{Miller}). A mass-transfer-rate of $2 \times 10^{-10}$ $M_{\odot} \over y$ would create an accretion disc around the gainer of TT Hya of 1,350 K, leaving only 400 K at its edge. The observed spot temperature of 17000 K would only be achieved by an extremely small sized spot characterized by an impossible value of $K$ ($\approx$ $10^{-7}$). If at the other hand the mass-transfer-rate would be as high as  $\approx$ $2 \times 10^{-7}$ $M_{\odot} \over y$ (as is the case for the similar system SW Cyg) one would obtain a disc temperature of $\approx$ 7,000 K leaving still 2,500 K at its edge. The value of $K$ would then have been reduced to a somewhat more acceptable value of $\approx$ $10^{-4}$. This value will, however, not be taken into account for our calibration because the data on $dM \over dt$ (the quantity $A$) and $T_{spot}$ (the quantity $K$) seem to contradict one another.

\subsubsection{V356~Sgr}

Stellar parameters are from Polidan (\cite{Polidan}), Simon (\cite{Simon}) and Peters \& Polidan (\cite{Peters}). From the observed time series of $dP \over dt$ Polidan (\cite{Polidan}) determines a mass-transfer-rate of $4 \times 10^{-7}$ $M_{\odot} \over y$. The gainer rotates at 6.3 times the synchronous velocity (Simon, \cite{Simon}). $L_{add}$ has been measured by Peters \& Polidan (\cite{Peters}) who also notice that a significant fraction of the circumstellar material is located near the surface of the gainer. This binary is in its era of mass-loss from the system so that the mass-transfer-rate exceeds its critical value given by relation (\ref{Mdotdisc}), leading to a value of the quantity $K$ of at least $1.9 \times 10^{-4}$.

\subsubsection{$\beta$ Lyr}

Harmanec \& Scholz (\cite{Harmanec2}) determine values of ${dP\over dt}$ and ${dM\over dt}$ which are very high (${dM\over dt}=3.4\times 10^{-5}$) $M_{\odot} \over y$ and are confirmed by Ak et al. (\cite{Ak}). This value is slightly larger than the one used by Simon (\cite{Simon}). $\beta$ Lyr is a binary with $T_{eff,g}$=28,000 K, $T_{disc}$=8,000 K and $T_{spot}$=20,000 K (Harmanec, \cite{Harmanec}). The best model is a thick Keplerian accretion disc around the gainer in which the effective gravitation at the edge is $\approx$ 0. Evaluated at the edge of the accretion disc we find $T_{edge}$ = 7,000 K, the accretion luminosity (428  $L_{\odot}$) and K ($\approx$ 0.012) evaluated with relation (\ref{Knewest}). Harmanec (\cite{Harmanec}) notices the existence of bipolar jets with wind velocities comparable to wind velocities of O stars. $\beta$ Lyr is blowing mass into the interstellar medium because its mass-transfer-rate is above the limit given by relation (\ref{Mdotdisc}) which is valid for a hot spot on the edge of an accretion disc. In that case the quantity $K$ needs to be $\approx$ 0.0066 so that (${dM\over dt}=3.4\times 10^{-5}$ $M_{\odot} \over y$) equals the critical value given by relation (\ref{Mdotdisc}). For the calibration of the quantity $K$ we have taken the last value which would also have been obtained with $T_{spot}$ = 23,000 K a number that differs not significantly from the value of 20,000 as K quoted by Harmanec (\cite{Harmanec}).

\subsection{$K$ as a function of total mass}

In this section we have determined values of $K$ for eight systems undergoing direct impact and three systems with accretion discs. No particular trend is found for the quantity $K$ as a function of many stellar parameters. However, one finds that low-mass binaries have larger values of $K$ than more massive ones. In order to include the quantity $K$ in the $Brussels~binary~evolutionary~code$ we have used the following tentative best fit relation:

\begin{equation}
{K=0.228 \times  \lbrack M_{d}Ê+ M_{g}  \rbrack ^ {-1.735}}
\label{K(M)} 
\end{equation} 

This relation has to be considered as very provisional. A similar but more accurate relation will, however, only be obtained if more precise and a greater number of observations on $L_{add}$, $S_{spot}$ and especially $T_{spot}$ will be available.

\section{Results}
\label{sec_Results}

A binary will lose mass only when the mass-transfer-rate rises above the critical rate as given by relation (\ref{FinalMdot}) for a direct impact system or relation (\ref{Mdotdisc}) if the the hot spot is located at the edge of an accretion disc. A large mass-transfer-rate will thus be the major driving mechanism of mass-loss from the system. A sufficiently large rate is achieved during a short era of fast mass-transfer soon after the onset of RLOF. Such high mass-transfer-rates are predicted by the binary evolution codes of:

\begin{flushleft}
$\bullet$~~Paczy\~{n}ski\\ 
\vspace{0.2cm}
$\bullet$~~Eggleton\\ 
\vspace{0.2cm}
$\bullet$~~$Brussels$ which is used in this analysis\\
\end{flushleft}

Paczy\~{n}ski (\cite{Paczynski}) quotes a maximum value of ${dM\over dt}$ as high as 3.4~$\times~10^{-3}$ $M_{\odot} \over y$ for a (16+10.67) $M_{\odot}$ binary with an initial period of 5.55 days. This value is almost recovered by the $Brussels$ code in its $conservative$ mode (1.8~$\times~10^{-3}$ $M_{\odot} \over y$). According to Paczy\~{n}ski (\cite{Paczynski}) this system transfers $\approx$ 9.1  $M_{\odot}$ from donor to gainer in 4,000 years only, during hydrogen shell burning of the donor. This amount of mass will certainly be too large for the gainer to be captured completely. Using the Eggleton code (\cite{Eggleton0} \& \cite{Eggleton2}), Yungelson  (\cite{Yungelson}) finds a mass-transfer-rate up to 5~$\times~10^{-4}$ $M_{\odot} \over y$ soon after the onset of RLOF during the hydrogen core burning of the donor for a (9+5.4) $M_{\odot}$ system with an initial period of 3 days. Also this mass-transfer-rate is almost predicted during the era of fast mass-transfer in this system by the $Brussels$ code in its $conservative$ mode (2~$\times~10^{-4}$ $M_{\odot} \over y$).

Low-mass binaries hardly achieve the requirements for significant mass-loss from the system. In this section we report the evolution of an initial binary of (9+5.4) $M_{\odot}$ with initial periods ranging from 1.4 to 5 days. We have calculated the evolution using the weak as well as the strong tidal interaction as outlined in section \ref{sec_Liberal}. The quantity $\beta$, defined by relation (\ref{eqn:beta}), was frequently treated as a free parameter. In this study  we calculate $\beta$ throughout the entire evolution of the binary.

\begin{figure*}[!ht]
\centering
\includegraphics[width=9.6cm]{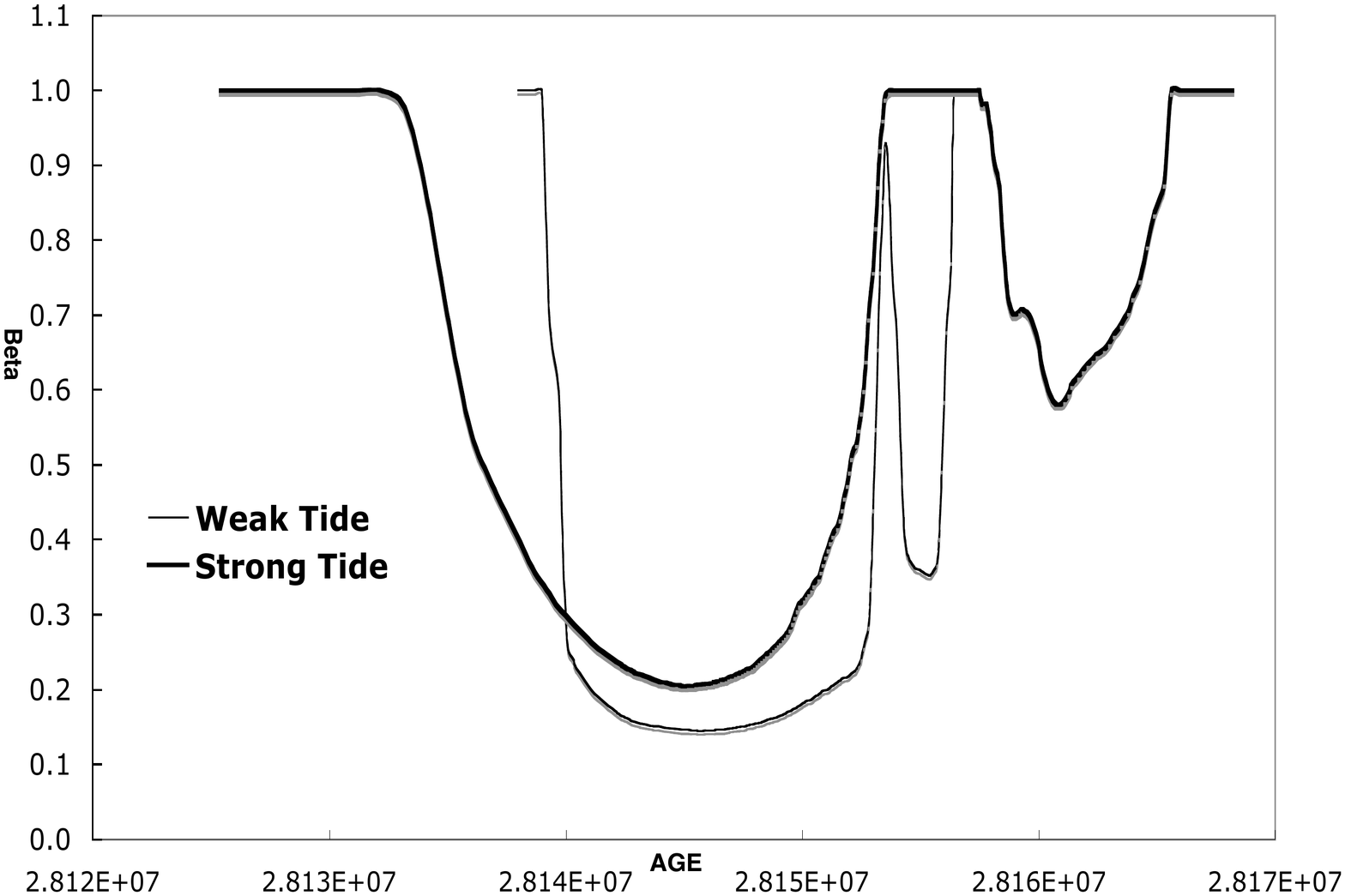}
\caption{$\beta$ as a function of time in and around the epoch of $liberal$ ($\beta$~$<$~1) evolution, for a system of (9+5.4) $M_{\odot}$, $P_{0}$~=~3.2 d, for the $weak$ and $strong$ tide approximation.} 
\label{fig_fig6}
\end{figure*}

Figure \ref{fig_fig6} shows the behavior of $\beta$ as indicator for the $liberal$ evolution of a (9+5.4) $M_{\odot}$ binary with an initial period 3.2 days. The two different tidal assumptions make hardly any difference for the result. The binary will almost always evolve $conservatively$, but in a short and violent era soon after RLOF ignition, mass will be lost from the system: 2.54 $M_{\odot}$ adapting the $strong~tidal$ interaction and 2.74 $M_{\odot}$ when the tidal interaction is $weak$. The small difference between the two results can be understood from relation (\ref{FinalMdot}). The critical mass-transfer-rate is indeed lowered by a high equatorial velocity of the gainer, which is more easily achieved with weak tides. However, the same critical rate is increased by higher rates of rotational kinetic energy of the gainer, which is favored by weak tidal interaction. In many cases both effects more or less cancel each other. Notice in Figure \ref{fig_fig6} that the system frequently undergoes $two$ separate stages of mass-loss ($\beta \ll$  1). The violent epoch of mass-loss indeed starts with $M_{d} > M_{g}$. The orbital period will shrink until $M_{d} = M_{g}$, which makes the impact parameter  $d$ (Figure \ref{fig_fig4}) so small that the accretion luminosity as calculated with expression (\ref{acclum}) will drop below the critical value. As soon as the system has acquired Algol characteristics with $M_{d} < M_{g}$, the orbit widens and a $liberal$ era can restart as soon as the mass-transfer-rate is sufficiently large. It happens that binaries run into superficial contact when during evolution $M_{d} \approx M_{g}$ so that the impact parameter vanishes. This leads to a brief pause in spin-up and further construction of the hot spot on the gainer. After some time, when $ M_{g}$ exceeds $M_{d}$ the contact is broken, as suggested for the short period binary CN And by Van Hamme et al. (\cite{Vanhamme}).

Table 1 covers a wide range of initial periods for (9+5.4) $M_{\odot}$ binary at birth. Mass-loss from the system always occurs soon after RLOF ignition. At the beginning of this violent epoch, the binary does not meet the criterion of Peters (\cite{Peters1}) stating that in a semi-detached binary, the donor needs to be less massive, cooler, fainter and larger than the gainer  before the binary can be labeled as Algol. At the end of the $liberal$ era the binary starts its life as an Algol. It is straightforward to determine the times that the Algol spends with changing values of mass-ratio and orbital period. 

\begin{table*}
\begin{center}
\begin{tabular}{cccccccccc} \hline
& P=1.8 & P=2 & P=2.25 & P=2.5 & P=2.75 & P=3  & P=3.2 & P=3.6 & P=4  \\ \hline
&  &  &  & strong tides &  &  &  &  &     \\
\hline
${\Delta~M}$ & 0.236 & 0.514 & 0.889 & 1.260 & 1.811 & 2.146 & 2.540 & 3.650 & 4.011    \\
${\Delta~t}_{liberal}$ & 22,300 & 20,900 & 29,400 & 31,500 & 32,500 & 29,700 & 38,000 & 32,200 & 31,300    \\
${\Delta~t}_{Algol}$ & 9,858,800 & 12,169,600 & 13,817,500 & 11,004,500 & 8,949,000 & 6,946,200 & 5,746,800 & 3,736,700 & 27,800    \\
\hline
&  &  &  & weak tides &  &  &  &  &     \\
\hline
${\Delta~M}$ & 0.084 & 0.023 & 0.152 & 0.933 & 1.545 & 2.391 & 2.737 & 3.315 & 3.237    \\
${\Delta~t}_{liberal}$ & 8,100 & 10,500 & 9,600 & 11,700 & 13,600 & 19,300 & 17,100 & 18,700 & 14,700    \\
${\Delta~t}_{Algol}$ & 10,067,000 & 11,887,600 & 14,570,400 & 10,970,200 & 8,778,700  & 7,091,300 & 5,970,500 & 3,670,700 & 1,412,600    \\
\hline
\end{tabular}
\caption{Characteristics of $liberal$ evolution of a binary (9+5.4) $M_{\odot}$ undergoing RLOF-A. Systems with periods below 1.8 days evolve almost $conservatively$ and are not mentioned. The mass-loss from the system hardly depends on the strength of the tidal interaction. The mass-loss is expressed in solar masses. The duration of $liberal$ evolution and of the Algol phase is in years.}
\end{center}
\label{tab_tab1}
\end{table*}

From Table 1 we learn that short initial periods yield conservative evolution only. The quantity of mass lost from the system rises with increasing initial orbital period. Binaries that start their RLOF at the end of hydrogen core burning (case-A) or at the beginning of hydrogen shell burning of the donor (case-B) lose a large amount of mass without being an Algol system for a long time. After a long era of $Algolism$ during hydrogen core burning of the donor, RLOF and subsequent Algol status are also achieved during hydrogen shell burning. During this epoch the mass-transfer-rate is, however, never sufficiently large to trigger mass-loss from the system.

\section{Conclusions}

\begin{figure*}[!ht]
\centering
\includegraphics[width=9.6cm]{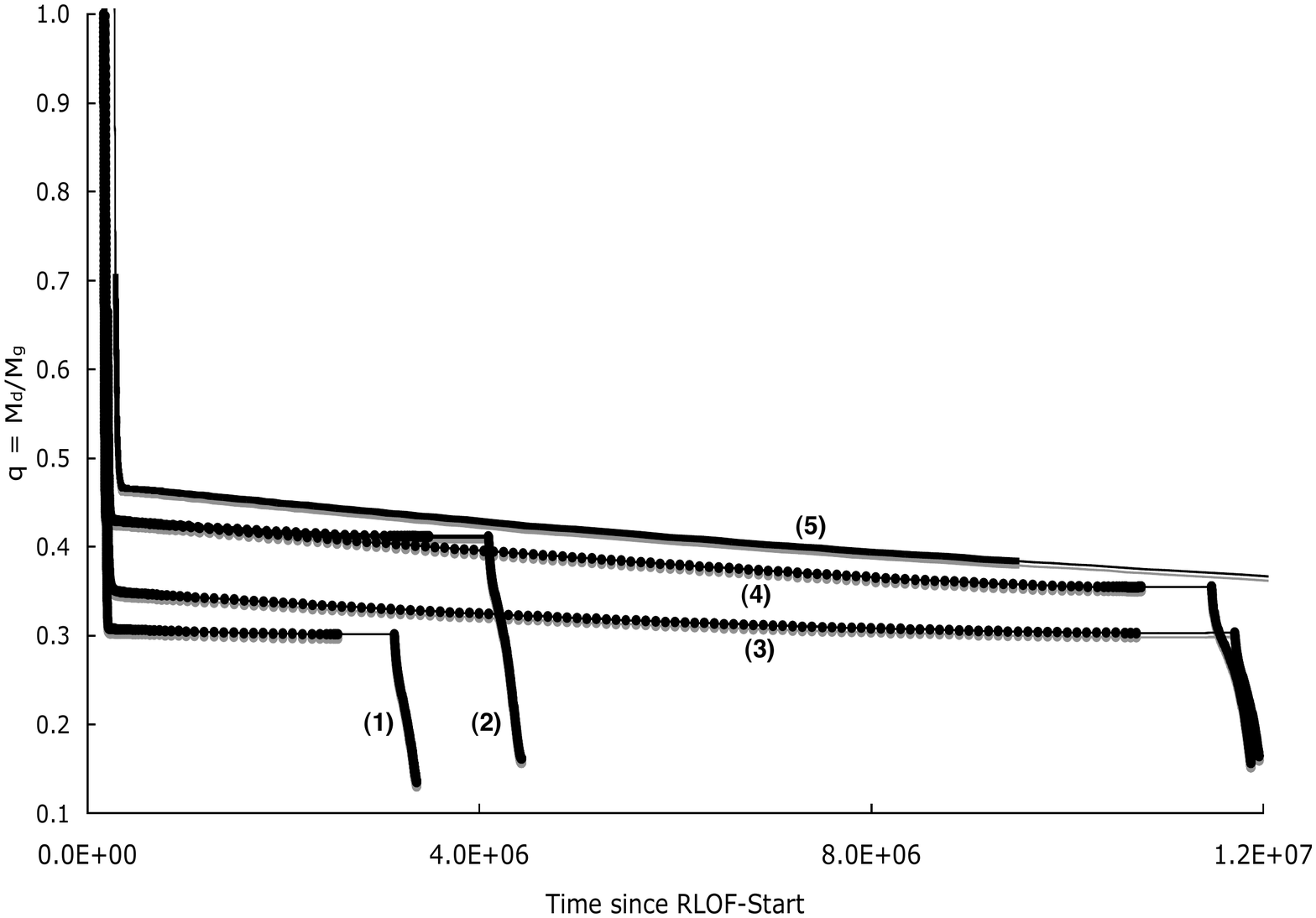}
\caption{Evolution of the mass-ratio $q$ = $M_{d}\over M_{g}$ with time after onset of RLOF. Binaries on the thin part of a curve are no Algols, unlike those located on the thick parts. Curves are labeled as follows: (1) Conservative evolution with $P_{0}$~=~3.6 d, (2) Liberal evolution with  $P_{0}$~=~3.6 d, (3) Conservative evolution with with $P_{0}$~=~2.5 d, (4) Liberal evolution with with $P_{0}$~=~2.5 d,  (5) The evolution with  $P_{0}$~=~1.6 d which is always conservative.} 
\label{fig_fig7}
\end{figure*}

Mass-loss into the interstellar medium is possible during the short stage of fast mass-transfer soon after RLOF ignition when the binary is not yet an Algol. Low-mass binaries hardly show sufficiently large mass-transfer-rates to reach the critical rate needed to overcome the binding energy of the system, as required by relation (\ref{FinalMdot}) for a direct impact system or (\ref{Mdotdisc}) if the direct impact is succeeded by the formation of a hot spot at the outer edge of a Keplerian accretion disc. Spin-up of the gainer and hot spots are frequently created but the joint energy of both mechanisms will overshoot the binding energy of the system mainly for intermediate-mass binaries. Therefore we have examined the evolution of the binary (9+5.4) $M_{\odot}$ with various initial orbital periods. We calculate mass-loss into the interstellar medium for binaries with an orbital period larger than 1.8 days. We have calculated the amount of mass lost with respectively strong and weak tidal interaction and found that the influence of the tides on the mass-loss is not significant. Figure  \ref{fig_fig7} shows the evolution of the mass-ratio $q$ = $M_{d}\over M_{g}$ with time since the start of RLOF. The shortest initial orbital period (P = 1.8 d) evolves almost $conservatively$ and yields a long living Algol that lives a very long time with $q$ $>$ 0.4. Table 1 indicates that the binary with an initial orbital period of 2.5 days loses $\approx$ 1 $M_{\odot}$ during its phase of rapid RLOF. The system shows Algol characteristics for more than 10 million years and $liberal$ evolution yields $q$-values which are systemically larger than those obtained with $conservative$ evolution ($\Delta~q$ a little below 0.1). A binary with an initial orbital period of 3.6 days loses more than 3 $M_{\odot}$ during its phase of rapid RLOF, shows its Algol characteristics for $\approx$ 4 million years and $liberal$ evolution yields a $\Delta~q$ a little above 0.1. Since we limited ourselves in this paper to binaries undergoing RLOF-A at first, we plan to examine binaries with larger initial orbital periods so that the fast phase of RLOF-B occurs towards a system having an accretion disc.

For our small sample of binaries with an initial primary mass of 9 $M_{\odot}$  and $q$~=~0.6, we used the initial orbital period distribution of Popova et al. (\cite{Popova}) and compared the obtained mass-ratio distribution with the result from conservative evolution. It is shown that the $q$-bin [0.4-0.6] which was poorly populated through $conservative$ binary evolution (less than 30$\%$) is now represented much better ($\approx$60$\%$). It is thus clear that our $liberal$ scenario meets the observed $q$-distribution of Algols better than $conservative$ evolution. It is, however, doubtful that detailed overall correspondence between theory and observations will be obtained, because all the low-mass binaries evolve almost $conservatively$, despite the fact that they also develop some spin-up and not so bright hot spots. $Conservative$ evolution is, however, not the general rule for binaries and the more massive among them can lose a significant amount of mass into space. In a near future we will complete our website  (\verb"http://www.vub.ac.be/astrofys/") containing an atlas of $conservative$ evolutionary calculations with the results from the $liberal$ calculations as presented in this paper. Future observations of mass-transfer-rates, energy contents, surface areas and temperatures of hot spots will certainly enable the researcher to refine the evaluation of the quantity $K$ for which relation (\ref{K(M)}) is only a first attempt.

\begin{acknowledgements}
      We thank Dr. Lev Yungelson and Prof. E. Van den Heuvel for very valuable suggestions and comments. Part of this work was supported by FWO under grant G.0044.05 and by the Research \& Development Department of the Vrije Universiteit Brussel
      \end{acknowledgements}

\end{document}